\documentclass[aps,prd,superscriptaddress,12pt,showpacs,notitlepage]{revtex4}
\advance \topmargin by -\headheight
\advance \topmargin by -\headsep

\usepackage{amssymb}
\usepackage{graphicx}

\evensidemargin \oddsidemargin
\begin{document}

\topmargin -.6in

\def\rh{{\hat \rho}}
\def\alie{{\hat{\cal G}}}
\newcommand{\sect}[1]{\setcounter{equation}{0}\section{#1}}
\renewcommand{\theequation}{\thesection.\arabic{equation}}

 %%      MACROS.TEX
%               macros formatting and equations
\def\rf#1{(\ref{eq:#1})}
\def\lab#1{\label{eq:#1}}
\def\nonu{\nonumber}
\def\br{\begin{eqnarray}}
\def\er{\end{eqnarray}}
\def\be{\begin{equation}}
\def\ee{\end{equation}}
\def\eq{\!\!\!\! &=& \!\!\!\! }
\def\foot#1{\footnotemark\footnotetext{#1}}
\def\lb{\lbrack}
\def\rb{\rbrack}
\def\llangle{\left\langle}
\def\rrangle{\right\rangle}
\def\blangle{\Bigl\langle}
\def\brangle{\Bigr\rangle}
\def\llbrack{\left\lbrack}
\def\rrbrack{\right\rbrack}
\def\lcurl{\left\{}
\def\rcurl{\right\}}
\def\({\left(}
\def\){\right)}
\newcommand{\nit}{\noindent}
\newcommand{\ct}[1]{\cite{#1}}
\newcommand{\bi}[1]{\bibitem{#1}}
\def\lskip{\vskip\baselineskip\vskip-\parskip\noindent}
\relax

%                     common physics symbols
\def\tr{\mathop{\rm tr}}
\def\Tr{\mathop{\rm Tr}}
\def\v{\vert}
\def\bv{\bigm\vert}
\def\Bgv{\;\Bigg\vert}
\def\bgv{\bigg\vert}
\newcommand\partder[2]{{{\partial {#1}}\over{\partial {#2}}}}
\newcommand\funcder[2]{{{\delta {#1}}\over{\delta {#2}}}}
\newcommand\Bil[2]{\Bigl\langle {#1} \Bigg\vert {#2} \Bigr\rangle}  %% <.|.>
\newcommand\bil[2]{\left\langle {#1} \bigg\vert {#2} \right\rangle} %% <.|.>
\newcommand\me[2]{\left\langle {#1}\bv {#2} \right\rangle} %% <.|.>
\newcommand\sbr[2]{\left\lbrack\,{#1}\, ,\,{#2}\,\right\rbrack}
\newcommand\pbr[2]{\{\,{#1}\, ,\,{#2}\,\}}
\newcommand\pbbr[2]{\lcurl\,{#1}\, ,\,{#2}\,\rcurl}

\def\ket#1{\mid {#1} \rangle}
\def\bra#1{\langle {#1} \mid}
\newcommand{\braket}[2]{\langle {#1} \mid {#2}\rangle}
%
%                    math symbols
\def\a{\alpha}
\def\at{{\tilde A}^R}
\def\atc{{\tilde {\cal A}}^R}
\def\atcm#1{{\tilde {\cal A}}^{(R,#1)}}
\def\b{\beta}
\def\dc{{\cal D}}
\def\d{\delta}
\def\D{\Delta}
\def\eps{\epsilon}
\def\vareps{\varepsilon}
\def\g{\gamma}
\def\G{\Gamma}
\def\grad{\nabla}
\def\h{{1\over 2}}
\def\l{\lambda}
\def\L{\Lambda}
\def\m{\mu}
\def\n{\nu}
\def\o{\over}
\def\om{\omega}
\def\O{\Omega}
\def\p{\phi}
\def\P{\Phi}
\def\pa{\partial}
\def\pr{\prime}
\def\pt{{\tilde \Phi}}
\def\qs{Q_{\bf s}}
\def\ra{\rightarrow}
\def\s{\sigma}
\def\S{\Sigma}
\def\t{\tau}
\def\th{\theta}
\def\Th{\Theta}
\def\tpp{\Theta_{+}}
\def\tmm{\Theta_{-}}
\def\tpg{\Theta_{+}^{>}}
\def\tms{\Theta_{-}^{<}}
\def\tp0{\Theta_{+}^{(0)}}
\def\tm0{\Theta_{-}^{(0)}}
\def\ti{\tilde}
\def\wti{\widetilde}
\def\jc{J^C}
\def\bj{{\bar J}}
\def\sj{{\jmath}}
\def\bsj{{\bar \jmath}}
\def\bp{{\bar \p}}
\def\vp{\varphi}
\def\vt{{\tilde \varphi}}
\def\faa{Fa\'a di Bruno~}
\def\ca{{\cal A}}
\def\cb{{\cal B}}
\def\ce{{\cal E}}
\def\cg{{\cal G}}
\def\cgh{{\hat {\cal G}}}
\def\ch{{\cal H}}
\def\chh{{\hat {\cal H}}}
\def\cl{{\cal L}}
\def\cm{{\cal M}}
\def\cn{{\cal N}}
\newcommand\sumi[1]{\sum_{#1}^{\infty}}   %% summation till infinity
\newcommand\fourmat[4]{\left(\begin{array}{cc}  %%   2x2 matrix
{#1} & {#2} \\ {#3} & {#4} \end{array} \right)}

%
%%%                    macros for Lie algebras
\def\lie{{\cal G}}
\def\kmlie{{\hat{\cal G}}}
\def\dlie{{\cal G}^{\ast}}
\def\elie{{\widetilde \lie}}
\def\edlie{{\elie}^{\ast}}
\def\hlie{{\cal H}}
\def\flie{{\cal F}}
\def\wlie{{\widetilde \lie}}
\def\f#1#2#3 {f^{#1#2}_{#3}}
\def\winf{{\sf w_\infty}}
\def\win1{{\sf w_{1+\infty}}}
\def\hwinf{{\sf {\hat w}_{\infty}}}
\def\Winf{{\sf W_\infty}}
\def\Win1{{\sf W_{1+\infty}}}
\def\hWinf{{\sf {\hat W}_{\infty}}}
\def\Rm#1#2{r(\vec{#1},\vec{#2})}          % r-operator kernel
\def\OR#1{{\cal O}(R_{#1})}           % R-coadjoint orbit
\def\ORti{{\cal O}({\widetilde R})}           % R-tilde-coadjoint orbit
\def\AdR#1{Ad_{R_{#1}}}              % R-adjoint group action
\def\dAdR#1{Ad_{R_{#1}^{\ast}}}      % R-coadjoint group action
\def\adR#1{ad_{R_{#1}^{\ast}}}       % R-coadjoint algebra action
\def\KP{${\rm \, KP\,}$}                 %% KP
\def\KPl{${\rm \,KP}_{\ell}\,$}         %% KP_l
\def\KPo{${\rm \,KP}_{\ell = 0}\,$}         %% KP_l=0
\def\mKPa{${\rm \,KP}_{\ell = 1}\,$}    %% modified KP-1
\def\mKPb{${\rm \,KP}_{\ell = 2}\,$}    %% modified KP-2
%
%       fake blackboard bold macros for reals, complex, etc.
\def\rlx{\relax\leavevmode}
\def\inbar{\vrule height1.5ex width.4pt depth0pt}
\def\IZ{\rlx\hbox{\sf Z\kern-.4em Z}}
\def\IR{\rlx\hbox{\rm I\kern-.18em R}}
\def\IC{\rlx\hbox{\,$\inbar\kern-.3em{\rm C}$}}
\def\IN{\rlx\hbox{\rm I\kern-.18em N}}
\def\IO{\rlx\hbox{\,$\inbar\kern-.3em{\rm O}$}}
\def\IP{\rlx\hbox{\rm I\kern-.18em P}}
\def\IQ{\rlx\hbox{\,$\inbar\kern-.3em{\rm Q}$}}
\def\IF{\rlx\hbox{\rm I\kern-.18em F}}
\def\IG{\rlx\hbox{\,$\inbar\kern-.3em{\rm G}$}}
\def\IH{\rlx\hbox{\rm I\kern-.18em H}}
\def\II{\rlx\hbox{\rm I\kern-.18em I}}
\def\IK{\rlx\hbox{\rm I\kern-.18em K}}
\def\IL{\rlx\hbox{\rm I\kern-.18em L}}
\def\one{\hbox{{1}\kern-.25em\hbox{l}}}
\def\0#1{\relax\ifmmode\mathaccent"7017{#1}%
B        \else\accent23#1\relax\fi}
\def\omz{\0 \omega}
%
%               \ltimes=\semiproduct
\def\ltimes{\mathrel{\vrule height1ex}\joinrel\mathrel\times}
\def\rtimes{\mathrel\times\joinrel\mathrel{\vrule height1ex}}
%
%               This defines remark, proposition etc.
\def\mark{\noindent{\bf Remark.}\quad}
\def\prop{\noindent{\bf Proposition.}\quad}
\def\theor{\noindent{\bf Theorem.}\quad}
\def\name{\noindent{\bf Definition.}\quad}
\def\exam{\noindent{\bf Example.}\quad}
\def\proof{\noindent{\bf Proof.}\quad}
%%
%
%       This defines the journal citations
%
\def\PRL#1#2#3{{\sl Phys. Rev. Lett.} {\bf#1} (#2) #3}
\def\NPB#1#2#3{{\sl Nucl. Phys.} {\bf B#1} (#2) #3}
\def\NPBFS#1#2#3#4{{\sl Nucl. Phys.} {\bf B#2} [FS#1] (#3) #4}
\def\CMP#1#2#3{{\sl Commun. Math. Phys.} {\bf #1} (#2) #3}
\def\PRD#1#2#3{{\sl Phys. Rev.} {\bf D#1} (#2) #3}
\def\PLA#1#2#3{{\sl Phys. Lett.} {\bf #1A} (#2) #3}
\def\PLB#1#2#3{{\sl Phys. Lett.} {\bf #1B} (#2) #3}
\def\JMP#1#2#3{{\sl J. Math. Phys.} {\bf #1} (#2) #3}
\def\PTP#1#2#3{{\sl Prog. Theor. Phys.} {\bf #1} (#2) #3}
\def\SPTP#1#2#3{{\sl Suppl. Prog. Theor. Phys.} {\bf #1} (#2) #3}
\def\AoP#1#2#3{{\sl Ann. of Phys.} {\bf #1} (#2) #3}
\def\PNAS#1#2#3{{\sl Proc. Natl. Acad. Sci. USA} {\bf #1} (#2) #3}
\def\RMP#1#2#3{{\sl Rev. Mod. Phys.} {\bf #1} (#2) #3}
\def\PR#1#2#3{{\sl Phys. Reports} {\bf #1} (#2) #3}
\def\AoM#1#2#3{{\sl Ann. of Math.} {\bf #1} (#2) #3}
\def\UMN#1#2#3{{\sl Usp. Mat. Nauk} {\bf #1} (#2) #3}
\def\FAP#1#2#3{{\sl Funkt. Anal. Prilozheniya} {\bf #1} (#2) #3}
\def\FAaIA#1#2#3{{\sl Functional Analysis and Its Application} {\bf #1} (#2)
#3}
\def\BAMS#1#2#3{{\sl Bull. Am. Math. Soc.} {\bf #1} (#2) #3}
\def\TAMS#1#2#3{{\sl Trans. Am. Math. Soc.} {\bf #1} (#2) #3}
\def\InvM#1#2#3{{\sl Invent. Math.} {\bf #1} (#2) #3}
\def\LMP#1#2#3{{\sl Letters in Math. Phys.} {\bf #1} (#2) #3}
\def\IJMPA#1#2#3{{\sl Int. J. Mod. Phys.} {\bf A#1} (#2) #3}
\def\AdM#1#2#3{{\sl Advances in Math.} {\bf #1} (#2) #3}
\def\RMaP#1#2#3{{\sl Reports on Math. Phys.} {\bf #1} (#2) #3}
\def\IJM#1#2#3{{\sl Ill. J. Math.} {\bf #1} (#2) #3}
\def\APP#1#2#3{{\sl Acta Phys. Polon.} {\bf #1} (#2) #3}
\def\TMP#1#2#3{{\sl Theor. Mat. Phys.} {\bf #1} (#2) #3}
\def\JPA#1#2#3{{\sl J. Physics} {\bf A#1} (#2) #3}
\def\JSM#1#2#3{{\sl J. Soviet Math.} {\bf #1} (#2) #3}
\def\MPLA#1#2#3{{\sl Mod. Phys. Lett.} {\bf A#1} (#2) #3}
\def\JETP#1#2#3{{\sl Sov. Phys. JETP} {\bf #1} (#2) #3}
\def\JETPL#1#2#3{{\sl  Sov. Phys. JETP Lett.} {\bf #1} (#2) #3}
\def\PHSA#1#2#3{{\sl Physica} {\bf A#1} (#2) #3}
\def\PHSD#1#2#3{{\sl Physica} {\bf D#1} (#2) #3}
\def\PJA#1#2#3{{\sl Proc. Japan. Acad} {\bf #1A} (#2) #3}
\def\JPSJ#1#2#3{{\sl J. Phys. Soc. Japan} {\bf #1} (#2) #3}
%%%

\vspace*{-1cm}

\vskip 2cm

\vspace{.2in}

\title{ Some aspects of self-duality and generalised BPS theories
 }

%%\vspace{.5cm}

\author{C. Adam}
\affiliation{Departamento de F\'isica de Part\'iculas, Universidad de Santiago de Compostela and Instituto Galego de F\'isica de Altas Enerxias (IGFAE) E-15782 Santiago de Compostela, Spain}
\author{L. A. Ferreira}
\affiliation{Instituto de F\'\i sica de S\~ao Carlos; IFSC/USP;
Universidade de S\~ao Paulo  \\ 
Caixa Postal 369, CEP 13560-970, S\~ao Carlos-SP, Brazil}
\author{E.  da Hora}
\affiliation{Departamento de F\'isica,
Universidade Federal de Maranh\~ao,\\ 65080-805, S\~ao Lu\'is, Maranh\~ao, Brazil}
\author{A. Wereszczynski}
\affiliation{Institute of Physics,  Jagiellonian University,
Reymonta 4, Krak\'{o}w, Poland}
\author{W. J. Zakrzewski}
\affiliation{Department of Mathematical Sciences,\\
 University of Durham, Durham DH1 3LE, U.K.}

\begin{abstract}
If a scalar field theory in (1+1) dimensions possesses soliton solutions obeying first order BPS equations, then, in general, it is possible to find an infinite number of related field theories with BPS solitons which obey closely related BPS equations. We point out that this fact may be understood as a simple consequence of an appropriately generalised notion of self-duality. 
We show that this self-duality framework enables us to generalize to higher
dimensions the construction of new solitons from already known solutions.
By performing simple field transformations our procedure allows us to relate
solitons with different topological properties.
We present several interesting examples of such solitons in two and three dimensions.

\end{abstract} 
\maketitle
\newpage

\section{Introduction}
\label{sec:intro}
\setcounter{equation}{0}
Many nonlinear field theories possess nontrivial static solutions of finite energy called solitons, which behave similarly to particles in several respects.  
Among these  especially interesting are the so-called topological solitons \cite{man-sut-book} which obey non-trivial boundary conditions. Topological solitons are stable because a deformation of the boundary conditions corresponding to the vacuum configuration would cost an infinite amount of energy and is, therefore, impossible. Topological solitons have a wide area of applications ranging from condensed matter systems to particle theory and cosmology. A powerful tool in the search for soliton solutions is provided by the so-called Bogomolnyi bounds \cite{bogo}, \cite{prasad}, that is, bounds on the soliton energies in terms of a topological charge or a homotopy invariant. In many cases it can be shown that topological solitons which saturate this bound must obey certain first order differential equations (BPS equations). The Bogomolnyi bound therefore both simplifies the task of finding solutions (first order BPS equations instead of second order Euler--Lagrange equations) and guarantees that the resulting BPS solutions are true (global) minima of the energy in the corresponding topological sector ({\it i.e.}, for the given boundary conditions). 

As a matter of fact, BPS equations frequently can be understood as self-duality equations, that is, the equality of two expressions (usually functions of the basic fields and their first derivatives) which are viewed as duals of each other in some sense. A slightly different point of view, which will be very useful for our purposes, may be found by reversing this logic. Thus we start with the self-duality equation $A=\widetilde A$ of two dual objects $A$ and $\widetilde A$ together with the assumption that the two objects combine into a homotopy invariant. The Bogomolnyi bounds and BPS equations for a related class of energy functionals are then the derived results, {\it i.e.},  consequences of the self-duality equations (see Section II for details).     

The simplest system possessing topological solitons, Bogomolnyi bounds and BPS equations is the theory of a scalar field in (1+1) dimensions
\be \lab{standard-lag}
{\cal L}=\frac{1}{2} \partial_\mu \varphi \partial^\mu \varphi - V(\varphi)
\ee
(our metric convention is $ds^2 = dt^2 - dx^2$) where $V(\varphi )\ge 0$, with the static energy functional
\be \lab{energy-1}
E[\varphi ] = \int dx \left( \frac{1}{2}\varphi '^2 +V \right) = \frac{1}{2} \int dx \left( \varphi ' \mp \sqrt{2V}\right)^2 \pm  
\int dx \varphi ' \sqrt{2V} 
\ee
(here $\varphi ' \equiv \frac{d}{d x} \varphi$). Finiteness of the energy requires that the potential $V$ has at least one vacuum value $\varphi = \varphi_1$, where $V(\varphi_1)=0$, such that field configurations $\varphi (x)$ may approach the vacuum for large $x$. If the potential has at least two vacua $\varphi_i$, then in general (that is, for sufficiently well behaved potentials) topological kink solutions interpolating between two adjacent vacua do exist. Assuming $\varphi_1 < \varphi_2$, the kink/antikink interpolating between $\varphi_1$ and $\varphi_2$ solves the first order BPS equations
\be \lab{BPS-eq-1}
\varphi' = \pm\sqrt{2V}
\ee
together with the boundary conditions 
\be
\lim_{x \to \mp\infty} \varphi (x) = \varphi_1 \; ,\quad \lim_{x\to \pm\infty} \varphi (x) = \varphi_2.
\ee
Introducing $W_\varphi  (\varphi ) =\sqrt{2V(\varphi)}$, the kink energy (the Bogomolnyi bound) is
\be
E = \int_{-\infty}^\infty  dx \varphi '  W_\varphi = \int_{\varphi_1}^{\varphi_2} d\varphi W_\varphi = W(\varphi_2) - W(\varphi_1)
\ee
and obviously only depends on the theory (the potential) and on the boundary conditions, but not on the field configuration. Thus, it is a homotopy invariant. 

As we shall discuss below, the BPS equation \rf{BPS-eq-1} has a simple behaviour under certain target space transformations. This fact led Bazeia and collaborators \cite{bazeia} to propose a procedure (``deformation") to generate infinite families of field theories with BPS kink solutions such that the kinks of this infinite family are related to the known kink solution of a given ``seed" theory just by changes of the field variable (coordinate transformations on target space). Putting these results into a more general context is one of the aims of the present work. As we are also interested in higher-dimensional generalisations, let us point out that the simple (BPS) scalar field theory \rf{standard-lag} has certain rather natural generalisations to BPS models in higher dimensions. The field theory \rf{standard-lag} itself, or a multi-field generalisation thereof, does not give rise to soliton solutions in higher dimensions as a consequence of Derrick's theorem. But a slight re-interpretation of the simple (1+1) dimensional Lagrangian \rf{standard-lag} permits us to find an expression amenable to higher-dimensional generalisations. Indeed, let us introduce the (off-shell conserved) ``topological current"
\be 
 j^\mu = \epsilon^{\mu\nu} \partial_\nu \varphi \; ,\quad \partial_\mu j^\mu \equiv 0
\ee
then the kinetic term may be written as $\partial^\mu \varphi \partial_\mu \varphi = - j^\mu j_\mu$ leading to the Lagrangian
\be
{\cal L}=-\frac{1}{2}j^\mu j_\mu - V.
\ee

This may be generalised easily to higher dimensions by introducing the appropriate topological currents. Let us consider a theory of $d$ real scalar fields $\varphi_a $, $a=1 ,\ldots ,d$ in $(d+1)$ space-time dimensions. Then a simple generalisation of the topological current takes the form
\be \lab{top-curr-1}
j^\mu = \epsilon^{\mu \mu_1 \cdots \mu_d} \partial_{\mu_1} \varphi_1 \ldots \partial_{\mu_d} \varphi_d 
\; ,\quad \partial_\mu j^\mu \equiv 0 .
\ee
For the resulting energy functional for static configurations in $d$-dimensional space, {\it i.e.} given by
\be
E[\varphi_a] = \int d^d x \left( \frac{1}{2}(j^0)^2 + V(\varphi_a) \right) ,
\ee
soliton solutions are not excluded by Derrick's theorem, because the two terms scale exactly oppositely under Derrick's scaling. And, indeed, field theories based on versions of the above energy functional \cite{tch} - \cite{baby-dual} are known to support both Bogomolnyi bounds and topological BPS solitons \cite{BPS-bS} - \cite{baby-dual}. Among these BPS theories there are BPS submodels of some well-known and relevant non-linear field theories like, {\it e.g.}, the Skyrme \cite{skyrme} and baby Skyrme \cite{baby}, \cite{holom} models or the abelian Higgs model, which makes them all the more interesting also from a phenomenological perspective (see, {\it e.g.}, \cite{BPS-Sk-app}). 

Given these results, one rather obvious question then arises - as to whether a version of the ``deformation procedure" mentioned above generalises to the higher-dimensional theories, and whether this procedure can be used also in these cases to find soliton solutions of many different theories starting from a soliton of a certain ``seed" theory.    

One of the aims of this paper is to  answer positively this question and to put the corresponding BPS theories into a more general context, starting from an appropriately generalised notion of self-duality. 

Concretely, in Section II we introduce the concept of self-duality in a rather general form and demonstrate that it leads directly to the first order BPS equations and Bogomolnyi bounds for a very large class of field theories (actually much larger than the class of theories discussed explicitly in this paper). In Section III, we introduce a class of generalised scalar field theories in (1+1) dimensions and show that they also fit into the self-duality framework. We then express the deformation procedure of Bazeia and collaborators in this framework and explain its geometrical underpinning. In Section IV, we introduce rather general classes of field theories in higher dimensions based on the topological current \rf{top-curr-1} which, again, naturally fit into the self-duality framework. We also discuss some important differences from the one-dimensional case, especially related to the much more involved target space geometries and topologies and to the much larger symmetry groups of these higher-dimensional theories.  
We also briefly discuss some examples in 2 dimensions. In Section V we present some examples of 3-dim BPS solitons with the topology of skyrmions and monopoles, respectively, and show how they are related via self-duality.

\section{Generalised self-duality }
\setcounter{equation}{0}
The concept of self-duality plays a key role in many areas of physics, helping to develop exact and non-perturbative methods. In field theory, self-duality conditions may be understood as the underlying cause for the existence of BPS equations, that is, first order (usually partial) differential equations with two striking features. First of all, their solutions are also solutions of the second order Euler-Lagrange differential equations, and they lead to the saturation of a bound on a functional which is usually an energy or an Euclidean action. The reason one has to perform one integration less to solve the equations of motion, does not come from the use of dynamically conserved charges. It comes from the invariance of a functional under smooth (homotopic) variations of the fields, {\it i.e.} a topological charge $Q$. Such a charge is given by  an integral formula as
\be
Q= \int d^d x\, A_{\alpha}\, {\widetilde A}_{\alpha},
\lab{primordialtopcharge}
\ee
where the integration is performed over a spatial (or space-time) manifold of dimension $d$, and the quantities $A_{\alpha}$  and ${\widetilde A}_{\alpha}$,  which are functions of the fields and their first derivatives, but not of higher order derivatives, are the ones to be considered dual to each other. The meaning of the index $\alpha$ depends upon the particular theory under consideration. By ``topological'' we mean that $Q$ is a homotopy invariant, that is, invariant under smooth variations of the fields, {\it i.e.}
\be
\delta Q=0 \qquad\qquad \qquad \mbox{\rm without the use of the eqs. of motion.}
\lab{selfdual1}
\ee

The self-duality equation corresponds to the equality
\be
A_{\alpha}=\pm {\widetilde A}_{\alpha}.
\lab{selfdualeq}
\ee
The conditions \rf{selfdual1} and \rf{selfdualeq} imply the Euler-Lagrange equations corresponding to the extrema of the functional 
\be
S= \frac{1}{2}\,\int d^n x\, \left[A_{\alpha}^2+{\widetilde A}_{\alpha}^2\right]
\lab{action}
\ee
where $n$ does not necessarily have to be equal to $d$. To see this 
let us denote by $\vp_a$ the fields of such a theory, which for the moment may be scalars, spinors, vectors, {\it etc}.  Then, under smooth infinitesimal variations of the fields we have 
\br
\delta Q&=& 0 = \int d^d x\, \left[ \delta A_{\alpha}\, {\widetilde A}_{\alpha}+A_{\alpha}\, \delta {\widetilde A}_{\alpha}\right]
\nonumber\\
&=& \int d^d x\, \left[ {\widetilde A}_{\alpha}\frac{\delta A_{\alpha}}{\delta \vp_j}\,\delta \vp_j
+ {\widetilde A}_{\alpha}\frac{\delta A_{\alpha}}{\delta \partial_{\mu}\vp_j}\,\delta \partial_{\mu}\vp_j
+ A_{\alpha}\leftrightarrow {\widetilde A}_{\alpha}\right]
\nonumber\\
&=& \int d^d x\, \left[ {\widetilde A}_{\alpha}\frac{\delta A_{\alpha}}{\delta \vp_j}
-  \partial_{\mu}\({\widetilde A}_{\alpha}\frac{\delta A_{\alpha}}{\delta \partial_{\mu}\vp_j}\)
+A_{\alpha}\frac{\delta {\widetilde A}_{\alpha}}{\delta \vp_j}
-  \partial_{\mu}\(A_{\alpha}\frac{\delta {\widetilde A}_{\alpha}}{\delta \partial_{\mu}\vp_j}\)
\right]\delta\vp_j
\nonumber\\
&+& \int d^d x\, \partial_{\mu}\left[
{\widetilde A}_{\alpha}\frac{\delta A_{\alpha}}{\delta \partial_{\mu}\vp_j}\,\delta \vp_j +
A_{\alpha}\frac{\delta {\widetilde A}_{\alpha}}{\delta \partial_{\mu}\vp_j}\,\delta \vp_j \right].
\er
The last quantity is a surface term and it  vanishes if one requires, for instance, that the variations of the field at the boundary vanish. Thus, since $Q$ is invariant under arbitrary smooth variations of the fields (homotopic), we see that we must have the following relations between  $A_{\alpha}$ and ${\widetilde A}_{\alpha}$
\be
{\widetilde A}_{\alpha}\frac{\delta A_{\alpha}}{\delta \vp_j}
-  \partial_{\mu}\({\widetilde A}_{\alpha}\frac{\delta A_{\alpha}}{\delta \partial_{\mu}\vp_j}\)
+A_{\alpha}\frac{\delta {\widetilde A}_{\alpha}}{\delta \vp_j}
-  \partial_{\mu}\(A_{\alpha}\frac{\delta {\widetilde A}_{\alpha}}{\delta \partial_{\mu}\vp_j}\)=0 .
\lab{topological}
\ee
On the other hand, the Euler-Lagrange equations following from the functional  \rf{action} are given by 
\be
A_{\alpha}\frac{\delta A_{\alpha}}{\delta \vp_j}
-  \partial_{\mu}\(A_{\alpha}\frac{\delta A_{\alpha}}{\delta \partial_{\mu}\vp_j}\)
+{\widetilde A}_{\alpha}\frac{\delta {\widetilde A}_{\alpha}}{\delta \vp_j}
-  \partial_{\mu}\({\widetilde A}_{\alpha}\frac{\delta {\widetilde A}_{\alpha}}{\delta \partial_{\mu}\vp_j}\)=0.
\lab{euler}
\ee
So, clearly,  \rf{selfdualeq} and \rf{topological}  imply \rf{euler}. 

In the cases where the functional \rf{action} is positive, and when the dimensions $n$ and $d$ are equal, the self-duality equations \rf{selfdualeq} imply the satuaration of a useful bound. Indeed, one can write 
\be
S= \frac{1}{2}\,\int d^d x\, \left[A_{\alpha}\mp {\widetilde A}_{\alpha} \right]^2  \pm  Q \qquad \rightarrow \qquad S \geq \mid Q\mid
\lab{bound}
\ee
and the bound is achieved for self-dual configurations, {\it i.e.}, the solutions of \rf{selfdualeq}.

A prototype of a self-dual theory is a Yang-Mills system. In this case, one has the identifications $A_{\alpha}\equiv F_{\mu\nu}$ and ${\widetilde A}_{\alpha}\equiv {\widetilde F}_{\mu\nu}$, with ${\widetilde F}_{\mu\nu}=\frac{1}{2}\varepsilon_{\mu\nu\rho\sigma} \, F^{\rho\sigma}$ being the Hodge dual of the field tensor $ F_{\mu\nu}$. The relevant topological charge is  the Pontryagin number
\be
Q_{YM}=\int d^4x\, {\rm Tr}\(F_{\mu\nu}\,{\widetilde F}_{\mu\nu}\)
\ee
and the functional \rf{action} is the Yang-Mills Euclidean action
\be
S_{YM}=\frac{1}{4}\int d^4x\, {\rm Tr}\(F_{\mu\nu}\,F_{\mu\nu}\)=\frac{1}{8}\int d^4x\, \left[{\rm Tr}\(F_{\mu\nu}\,F_{\mu\nu}\)+
{\rm Tr}\({\widetilde F}_{\mu\nu}\,{\widetilde F}_{\mu\nu}\)\right] .
\ee
An even simpler example is provided by {\it e.g.}, the CP(1) model which may be parametrised by a complex field $u=\varphi_1 + i \varphi_2$ taking values in the
one-point compactified complex plane. Its energy functional is
\be
E = \int d^2 x \frac{\partial_j u  \partial_j\bar u}{(1+u\bar u)^2} = \int d^2 x \frac{(\partial_j \varphi_1)^2 + (\partial_j \varphi_2)^2}{(1+\vec \varphi^2)^2} ,
\ee
and the two dual objects may be chosen, {\it e.g.}, as
\be
A_\alpha = \frac{\partial_j \varphi_1}{1+\vec \varphi^2 } \; ,\quad \widetilde A_\alpha = 
\epsilon_{jk}\frac{\partial_k \varphi_2}{1+\vec \varphi^2 }
\ee
leading to the topological charge
\be 
Q_{CP(1)} = \frac{i}{2}\epsilon_{jk} \int d^2 x \frac{\partial_j u \partial_k \bar u}{(1+u\bar u)^2} =\pi k
\ee
where $k\in \mathbb{Z}$ is the degree (winding number) of the map induced by $u$. Finally, the self-duality equations are given by
\be
\partial_j \varphi_1 = \epsilon_{jk}\partial_k\varphi_2,
\ee
which are easily recognized as the Cauchy-Riemann equations.

Let us end this section with two comments which will be useful later on. Firstly, the topological charge \rf{primordialtopcharge} is obviously invariant under the simultaneous transformations $A_\alpha \to gA_\alpha$ and $\tilde A_\alpha \to g^{-1} \tilde A_\alpha$ where $g$ is a (for the moment arbitrary) function. This leads to the new self-duality equations and the action functional
\be
A_{\alpha}=\pm g^{-2}{\widetilde A}_{\alpha}
%%\lab{selfdualeq}
\ee
\be
S= \frac{1}{2}\,\int d^n x\, \left[g^2 A_{\alpha}^2+g^{-2}{\widetilde A}_{\alpha}^2\right] .
%%\lab{action}
\ee
Secondly, the self-duality equations are invariant under the simultaneous transformations $A_\alpha \to g A_\alpha$ and $\tilde A_\alpha \to g \tilde A_\alpha$. What is not obvious in this case is whether the resulting ``topological charge"
\be
Q_g = \int d^d x g^2 A_\alpha \widetilde A_\alpha
\ee
is still a homotopy invariant. In many cases, and for appropriate choices of the function $g$, this is the case. In these cases the corresponding energy or the action functional takes the form
\be
S= \frac{1}{2}\,\int d^n x\, g^2 \left[ A_{\alpha}^2+ {\widetilde A}_{\alpha}^2\right] .
%%\lab{action}
\ee

\section{One dimension}
\setcounter{equation}{0}
Let us now apply the ideas discussed above to the simple case of a scalar field $\vp$ in $(1+1)$ dimensions. We assume for the moment that the scalar field is restricted to take fixed values at spatial infinity ({\it e.g.} by the condition of finite energy),{\it i.e.}, $\varphi (x=-\infty ) = \vp_1$, $\vp (x=\infty )= \vp_2$. Then the simplest choice for a topological charge in such theories is  
\be
Q_1=\int_{-\infty}^{\infty} dx \, \frac{d\,\vp}{d\,x}= \vp\(\infty\)-\vp\(-\infty\) = \vp_2 - \vp_1 .
\ee

Clearly, smooth variations of the field which respect the boundary conditions leave this charge invariant. However, the same is true if one replaces the field by a function of it (as long as it respects ``similar" boundary conditions)
\br
Q_{1,\Phi} &=& \int_{-\infty}^{\infty} dx \, \frac{d\,\Phi\(\vp\)}{d\,x}= \int_{-\infty}^{\infty} dx \, \frac{ \pa \,\Phi\(\vp\)}{ \pa \, \vp}\, \frac{d\,\vp}{d\,x} \nonumber \\
&=& \int_{\vp_1}^{\vp_2} d\vp \frac{\pa \Phi}{\pa \vp} = \int_{\Phi_1}^{\Phi_2} d\Phi= \Phi_2 -\Phi_1 \equiv \Phi (\vp_2) - \Phi (\vp_1).
\er
If $\Phi (\vp)$ is invertible, then $\vp \to \Phi (\vp)$ can be interpreted as a coordinate transformation on the target space.

If we now make the identifications
\be
A_{\alpha}\equiv \frac{d\,\vp}{d\,x},\qquad\qquad\qquad {\widetilde A}_{\alpha}\equiv \frac{\pa \,\Phi\(\vp\)}{ \pa \, \vp}\equiv \sqrt{2\,V\(\vp\)}
\ee
then the self-duality equation \rf{selfdualeq} and energy functional \rf{action} lead to
\be
 \frac{d\,\vp}{d\,x}=\pm \sqrt{2\,V\(\vp\)}
 \lab{selfdual1d}
 \ee
and
 \be
 E=\int_{-\infty}^{\infty} dx \,\left[ \frac{1}{2}\, \( \frac{d\,\vp}{d\,x}\)^2+V\(\vp\)\right],
 \lab{energy1d}
 \ee
which coincide with the BPS equation \rf{BPS-eq-1} and the energy \rf{energy-1} of a scalar field theory in (1+1) dimensions. 

If we only know the $\vp$ derivative $G(\vp) = \pa \Phi /\pa \vp$, or if $G$ cannot be integrated globally, it is still true that the functional
\be
Q_{1,G}=\int_{-\infty}^{\infty} dx \, G\(\vp\)\frac{d\,\vp}{d\,x}
\lab{nice1dcharge}
\ee 
defines a homotopy invariant, {\it i.e.}, is an invariant under arbitrary smooth variations of the field. Indeed, one has 
\be
\delta\,Q_{1,G}=\int_{-\infty}^{\infty} dx \,\left[ \delta  \vp\,\frac{\pa \,G}{\pa \,\vp}\, \frac{d\,\vp}{d\,x}+ G\, \frac{d\, \delta\,\vp}{d\,x}\right] 
=\int_{-\infty}^{\infty} dx \,\delta\vp\, \left[\frac{ \pa \,G}{\pa \,\vp}\, \frac{d\,\vp}{d\,x}- \frac{d\,G}{d\,x} \right] +G\,\delta\,\vp\mid_{-\infty}^{\infty}
\ee
and so $\delta\,Q_{1,G}$ vanishes for variations of the field that vanish at infinity. If one makes the naive identifications $A_{\alpha}\equiv  \frac{d\,\vp}{d\,x}$ and ${\widetilde A}_{\alpha}\equiv G$, we are back at what we had before. However, let us take $G$ as $G\(\vp\)=g^2\(\vp\)\,\sqrt{2\,V\(\vp\)}$. Then \rf{nice1dcharge} becomes 
\be
Q_{1,G}=\int_{-\infty}^{\infty} dx \, \(g\(\vp\)\, \sqrt{2\,V\(\vp\)}\)\(g\(\vp\)\,\frac{d\,\vp}{d\,x}\).
\lab{nice1dcharge2}
\ee
If we now make the identifications
\be
A_{\alpha}\equiv g\(\vp\)\,\frac{d\,\vp}{d\,x},\qquad\qquad\qquad {\widetilde A}_{\alpha}\equiv g\(\vp\)\,\sqrt{2\,V\(\vp\)} 
\ee
we observe that this leads to the same self-duality equation, namely, \rf{selfdual1d}. However, the functional \rf{action} has now  become
\be
 E_G=\int_{-\infty}^{\infty} dx \,g\(\vp\)^2\,\left[ \frac{1}{2}\, \( \frac{d\,\vp}{d\,x}\)^2+V\(\vp\)\right] .
 \lab{energy1dnice}
 \ee

 The Euler-Lagrange equation following from the functional \rf{energy1dnice} is definitively different from the one  following from \rf{energy1d}. Thus, we conclude that the solutions of the self-duality equation \rf{selfdual1d} solve the Euler-Lagrange equations of an infinitely large family of theories parameterized by the functional $g\(\vp\)$. This is a remarkable fact.
 
 For the cases where $g$ is a derivative {\it i.e.}  $g\(\vp\)=\frac{\pa \,\phi\(\vp\)}{\pa \,\vp}$, and so can be interpreted as being a Jacobian of a change of variables, one finds that \rf{energy1dnice} now becomes
 \be
 E_G=\int_{-\infty}^{\infty} dx \,\left[ \frac{1}{2}\, \( \frac{d\,\phi}{d\,x}\)^2+{\bar V}\(\phi\)\right],
 \lab{energy1dnice2}
 \ee
 where we introduced the potential
 \be
 {\bar V}\(\phi\)= g\(\phi\)^2\, V\(\phi\) ,
 \ee
and we assumed that the relation between $\phi$ and $\vp$ is invertible, at least in the interval $\vp \in [\vp_1 ,\vp_2 ]$.

 Thus, self-dual solutions $\vp (x)$ of the theory  \rf{energy1d} can be mapped into self-dual solutions $\phi (x)$ of the theory \rf{energy1dnice2} by the simple map $\phi (x)= \phi (\vp (x))$. In fact, the self-duality equations for both theories are the same and given by \rf{selfdual1d}. This is the deformation procedure that Bazeia and collaborators have been using in applications of scalar field theories \cite{bazeia}, where $\phi (\vp)$ is called the ``deformation function". Despite its simplicity, this procedure has led to some useful applications. By extending $\phi (\varphi)$ to a non-invertible function on a larger interval one can, for example, relate theories with different numbers of vacua. Another possible application involves using an integrable model (like the sine Gordon model) as a ``seed" theory and in introducing parameter families of deformation functions, such that various consequences of small deformations away from integrability may be investigated \cite{fer-zak} (``quasi-integrability"). Very recently, the deformation procedure was employed to construct joint kink solutions of theories of various coupled scalar fields \cite{bazeia2} which, without this procedure, would have been a much more difficult task. Finally, the deformation may also be used to find families of BPS solutions for higher-dimensional field theories after dimensional reduction, {\em i.e.},  assuming a spherically symmetric ansatz for the fields, see, {\em e.g.}, 
\cite{casana}. 

Before ending this section, we want to make one more comment which will be useful for the higher-dimensional generalisations. Let us consider 
\be
A_\alpha = g(\vp ) \frac{d}{dx} \vp \; , \quad \widetilde A_\alpha = \sqrt{2V(\vp )}
\ee
leading to the energy functional
\be
  E_g=\int_{-\infty}^{\infty} dx \left[ \frac{1}{2}\, g\(\vp\)^2\, \( \frac{d\,\vp}{d\,x}\)^2+V\(\vp\)\right] 
 \lab{energy1dnice-tilde}
 \ee
where we assume, in addition, that $V$ has two zeros (vacua) at $\vp = \vp_1$ and $\vp = \vp_2$, and that $g>0$ in the interval $\vp \in [\vp_1 ,\vp_2 ]$. Then the corresponding self-duality equations imply that the kink solution interpolating between $\vp_1$ and $\vp_2$ only takes values in the finite interval (fundamental region) $\vp \in [\vp_1 ,\vp_2 ]$, that is, the target space manifold for this variational problem is given by this finite interval.  The topological charge then is given by
\br
Q_{1,g} &=& \int_{-\infty}^\infty dx g(\vp ) \; \sqrt{2V} \; \frac{d\vp}{dx} = \int_{\vp_1}^{\vp_2} d\vp g(\vp ) \sqrt{2V} \\
&=&   \int_{\cal M} d\Omega^{(1)}_g \sqrt{2V} = {\bf V}({\cal M}) \langle \sqrt{2V} \rangle_{\cal M} .
\lab{1d-top-charge}
\er
Here, the first line in this expression just tells us again that $Q_{1,g}$ can be expressed as a target space integral (does not depend on the configuration $\vp (x)$), as befits a topological charge. The second line introduces some differential geometric notation which will be useful later on. Concretely,  ${\cal M}$ is the target space manifold (the interval $[\vp_1 ,\vp_2]$) and ${\bf V}({\cal M})$ is its ``volume". Furthermore, the positive function $g(\vp )$ can be interpreted as a target space ``volume" density such that $d\Omega^{(1)}_g = gd\vp$ is the corresponding target space volume form. Finally, 
\be
\langle \sqrt{2V} \rangle = \frac{1}{{\bf V}({\cal M})} \int_{\cal M} d\Omega^{(1)}_g \sqrt{2V}
\ee
 is the average value of the target space function $\sqrt{2V}$ on the target space ${\cal M}$.

\section{Higher dimensions}
\setcounter{equation}{0}
 
\subsection{Topological charges }
Here we want to generalise these ideas to higher dimensions. Recalling our discussion in the introduction (the topological current
\rf{top-curr-1}), we start by considering the following class of topological charges 
\be
Q_d=\int d^dx \, K\(\vp_a, \partial_j \vp_a\) \; , \qquad
K\equiv B\(\vp_a\)\,\sum_{i_1\ldots i_d=1}^d \varepsilon_{i_1\ldots i_d} \partial_{i_1}\vp_1\ldots \partial_{i_d}\vp_d
\lab{topcharged}
\ee
in $(d+0)$ dimensions. They  generalise \rf{nice1dcharge} and involve a set of $d$ real scalar fields  $\vp_a$, $a=1,2,\ldots d$.
Here, $B\(\vp_a\)$ is an arbitrary function of the fields but not of their derivatives. 

Under smooth variations of the fields one finds that 
\be
\delta\,Q_d=\int d^dx \, \sum_{a=1}^d\delta\vp_a\left[-\sum_{j=1}^d \partial_j\(\frac{\pa K}{\pa \,\partial_j\vp_a}\)+ \frac{\pa \, K}{\pa \, \vp_a}\right]
\quad + \quad \hbox{\rm surface term}.
\ee
Thus  $Q_d$ is topological, {\it i.e.} is invariant under arbitrary smooth variations of the fields that vanish at spatial infinity, if $K$ satisfies the  equation  
\be
\sum_{j=1}^d \partial_j\(\frac{\pa K}{\pa \,\partial_j\vp_a}\)- \frac{\pa \, K}{\pa \, \vp_a}=0
\lab{nice0}
\ee 
for {\em arbitrary} field configurations.
In order to prove \rf{nice0}, we first note that 
\br
\frac{\pa K}{\pa \,\partial_j\vp_a}&=&B\(\vp\)\,
\sum_{i_1\ldots i_d=1}^d \delta_{j\,i_a}\varepsilon_{i_1\ldots i_d} \partial_{i_1}\vp_1\ldots \partial_{i_{a-1}}\vp_{a-1}\, \partial_{i_{a+1}}\vp_{a+1}\ldots \partial_{i_d}\vp_d
\lab{aaa}
\\
&=& B\(\vp\)\,
\sum_{i_1\ldots i_{a-1}\,i_{a+1}\ldots i_d=1}^d \varepsilon_{i_1\ldots i_{a-1}\, j\, i_{a+1} \ldots i_d} \partial_{i_1}\vp_1\ldots \partial_{i_{a-1}}\vp_{a-1}\, \partial_{i_{a+1}}\vp_{a+1}\ldots \partial_{i_d}\vp_d
\nonumber
\er
Note that when we act on \rf{aaa} with $\partial_j$ we get terms of the form $\partial_j\,\partial_{i_b} \vp_b$ contracted with the antisymmetric symbol $\varepsilon_{i_1\ldots i_{a-1}\, j\, i_{a+1} \ldots i_d} $, and so they vanish. In addition, we get also other terms where $\partial_j$ acts on $B\(\vp\)$. So we have
\br
\sum_{j=1}^d \partial_j\(\frac{\pa K}{\pa \,\partial_j\vp_a}\)&=&
\sum_{b=1}^d\frac{\pa \, B\(\vp\)}{\pa \, \vp_b} \times 
\nonumber\\
&\times&
\sum_{i_1\ldots i_{a-1}\,j\, i_{a+1}\ldots i_d=1}^d \partial_j\vp_b\,\varepsilon_{i_1\ldots i_{a-1}\, j\, i_{a+1} \ldots i_d} \partial_{i_1}\vp_1\ldots \partial_{i_{a-1}}\vp_{a-1}\, \partial_{i_{a+1}}\vp_{a+1}\ldots \partial_{i_d}\vp_d
\nonumber
\er
However, if $b$ is equal to one of the indices of the $\vp$'s under the other derivatives one gets zero since the two derivatives are contracted with the $\varepsilon$ symbol. So, $b$ must be equal to the index of the field which has disappeared, {\it i.e.} $b=a$, and we get
\br
\sum_{j=1}^d \partial_j\(\frac{\pa K}{\pa \,\partial_j\vp_a}\)&=& 
\frac{\pa \, B\(\vp\)}{\pa \, \vp_a} \, 
\sum_{i_1\ldots i_d=1}^d \varepsilon_{i_1\ldots i_d} \partial_{i_1}\vp_1\ldots \partial_{i_d}\vp_d
=\frac{\pa \, K}{\pa \, \vp_a}
\er
which demonstrates that we have proved \rf{nice0}.

Following the one dimensional case let us write $B$ as $B\(\vp_a\)= b\(\vp_a\)^2\, M\(\vp_a\)\,\sqrt{2\, V\(\vp_a\)}$, and make the identifications
\be
A_{\alpha}\equiv b\(\vp_a\)\,  M\(\vp_a\)\,\sum_{i_1\ldots i_d=1}^d \varepsilon_{i_1\ldots i_d} \partial_{i_1}\vp_1\ldots \partial_{i_d}\vp_d, \qquad\qquad\qquad
{\widetilde A}_{\alpha}\equiv b\(\vp_a\)\, \sqrt{2\, V\(\vp_a\)}
\ee
and so we note that $Q_d$ takes the form of \rf{primordialtopcharge} and, as we just have proved, it satisfies \rf{selfdual1}. In addition, the self-duality equation 
\rf{selfdualeq} now becomes
\be
 M\(\vp_a\)\,\sum_{i_1\ldots i_d=1}^d \varepsilon_{i_1\ldots i_d} \partial_{i_1}\vp_1\ldots \partial_{i_d}\vp_d=\pm \sqrt{2\, V\(\vp_a\)} .
\lab{selfdualeqd}
\ee
It then follows that the solutions of \rf{selfdualeqd} solve the Euler-Lagrange equations corresponding to the functional
\be
S= \frac{1}{2}\int d^d x\, \, b\(\vp_a\)^2\left[\( M\(\vp_a\)\,\sum_{i_1\ldots i_d=1}^d \varepsilon_{i_1\ldots i_d} \partial_{i_1}\vp_1\ldots \partial_{i_d}\vp_d\)^2 + V\(\vp_a\)\right] .
\lab{actiond}
\ee
So, we have in $d$ dimensions the same situation we had in the one-dimensional case. Solutions of the self-duality equations \rf{selfdualeqd} are solutions of an infinite set of theories, defined by the action/energy \rf{actiond}, and parameterized by the function $b\(\vp_a\)$. That is an even more remarkable fact.

One can now think of a generalization to $d$ dimensions of the deformation procedure of Bazeia and collaborators \cite{bazeia}. Consider a field theory defined by the functional \rf{actiond} with $b=1$, {\it i.e.},
\be
S_1= \frac{1}{2}\int d^d x\, \left[\( M\(\vp_a\)\,\sum_{i_1\ldots i_d=1}^d \varepsilon_{i_1\ldots i_d} \partial_{i_1}\vp_1\ldots \partial_{i_d}\vp_d\)^2 + V\(\vp_a\)\right].
\lab{actionds1}
\ee
Next, define the field transformation
\be
\vp_a=\vp_a\(\phi\).
\ee
Then
\be
\sum_{i_1\ldots i_d=1}^d \varepsilon_{i_1\ldots i_d} \partial_{i_1}\vp_1\ldots \partial_{i_d}\vp_d=
\sum_{i_1\ldots i_d=1}^d \varepsilon_{i_1\ldots i_d} \frac{\pa \vp_1}{\pa \phi_{a_1}} \partial_{i_1}\phi_{a_1}\ldots \frac{\pa \vp_d}{\pa \phi_{a_d}}\partial_{i_d}\phi_{a_d}. 
\ee
However,
\be
\sum_{i_1\ldots i_d=1}^d \varepsilon_{i_1\ldots i_d}\partial_{i_1}\phi_{a_1}\ldots \partial_{i_d}\phi_{a_d} = \varepsilon_{a_1\ldots a_d}
\sum_{i_1\ldots i_d=1}^d \varepsilon_{i_1\ldots i_d}\partial_{i_1}\phi_{1}\ldots \partial_{i_d}\phi_{d}
\ee
and so
\br
\sum_{i_1\ldots i_d=1}^d \varepsilon_{i_1\ldots i_d} \partial_{i_1}\vp_1\ldots \partial_{i_d}\vp_d&=&
\sum_{i_1\ldots i_d=1}^d\varepsilon_{a_1\ldots a_d}\frac{\pa \vp_1}{\pa \phi_{a_1}}\ldots 
\frac{\pa \vp_d}{\pa \phi_{a_d}}\,
\sum_{i_1\ldots i_d=1}^d \varepsilon_{i_1\ldots i_d}\partial_{i_1}\phi_{1}\ldots \partial_{i_d}\phi_{d}
\nonumber\\
&=&\mid \frac{\pa \vp}{\pa \phi}\mid\,
\sum_{i_1\ldots i_d=1}^d \varepsilon_{i_1\ldots i_d}\partial_{i_1}\phi_{1}\ldots \partial_{i_d}\phi_{d},
\er
where $\mid \frac{\pa \vp}{\pa \phi}\mid$ is the Jacobian of the transformation. Thus, if one chooses $b\(\vp_a\)$ to be given by 
\be
b\(\vp_a\)\equiv \mid \frac{\pa\vp}{\pa \phi}\mid^{-2}
\ee
one finds that the functional \rf{actiond} becomes
\be
S_2= \frac{1}{2}\int d^d x\, \left[\( M\(\phi_a\)\,\sum_{i_1\ldots i_d=1}^d \varepsilon_{i_1\ldots i_d} \partial_{i_1}\phi_1\ldots \partial_{i_d}\phi_d\)^2 + {\bar V}\(\phi_a\)\right],
\lab{actiond2}
\ee
where the new potential is defined as
\be
{\bar V}\equiv\mid \frac{\pa \vp}{\pa \phi}\mid^{-2}\, V.
\ee 
Thus, all solutions of the self-duality equation \rf{selfdualeqd}, which are solutions of \rf{actionds1}, are mapped into the self-dual solutions of the theory \rf{actiond2}.

\subsection{Target spaces and vacuum structure}

The topological charge \rf{topcharged} is a homotopy invariant by construction, but we did not discuss yet under which conditions it leads to a nontrivial (nonzero) energy bound or be related to genuine topological properties like, e.g., elements of homotopy groups (winding numbers, {\it etc.}). 
There are, in principle, many possibilities to equip field theories with nontrivial topological structures (see, {\it e.g.}, \cite{man-sut-book}), but here we shall restrict ourselves to the class of energy functionals 
\be
E = \frac{1}{2}\int d^d x\, \, \left[\( M\(\vp_a\)\,\sum_{i_1\ldots i_d=1}^d \varepsilon_{i_1\ldots i_d} \partial_{i_1}\vp_1\ldots \partial_{i_d}\vp_d\)^2 + V\(\vp_a\)\right] 
\lab{actiondagain}
\ee
for which topology enters via some conditions/restrictions on the two functions $M$ and $V$ (we remark that the functional \rf{actiond} may be rewritten as \rf{actiondagain} by a simple redefinition of $M$ and $V$). The condition of finite energy requires  $V$ to have at least one vacuum, {\it i.e.}, value $\vec \vp = \vec \vp_0$ such that $V(\vec \vp_0)=0$. If the vacuum of $V$ is just a point $\vec \vp_0 \in {\cal M}$ of the target space, then for finite energy the field vector $\vec \vp$ must approach this point in the limit of infinite distance $\vert \vec x \vert \to \infty$ independently of the direction of $\vec x$. As a result of this requirement  finite energy field configurations have as their true base space the one-point compactified Euclidean space $\mathbb{R}^d_0$ which is topologically equivalent to the $d$-dimensional sphere $\mathbb{S}^d$. Finite energy field configurations are, hence, maps from $\mathbb{S}^d $ to ${\cal M}$ and may be classified by the corresponding homotopy group $\pi_d ({\cal M})$. If the target space, too, has the topology of the sphere $\mathbb{S}^d$, then $\pi_d (\mathbb{S}^d)=\mathbb{Z}$, and the corresponding topological index is a winding number, as is the case, {\it e.g.}, of the Skyrme or baby Skyrme models.

Another possibility involves endowing the theory with nontrivial topology even for topologically trivial target spaces ${\cal M}$. This occurs when the potential assumes its vacuum value $V=0$ for fields $\vp_a$ taking their values in a nontrivial submanifold ${\cal V \in M}$, the vacuum manifold
\be
{\cal V} = \{ \vec \vp \in {\cal M}\; \vert \;  V(\vec \vp )=0 \} .
\ee
This happens, {\it e.g.}, in field theories with spontaneous symmetry breaking.
In such a case, the finite energy field configurations $\vp_a (\vec x)$ do not have then to assume a unique value in the limit $\vert \vec x\vert \to \infty$ but, instead, may take different  values $\vec \vp \in {\cal V}$ in different directions of $\vec x$. They define, therefore, maps from the ``boundary" of the $d$-dimensional space (the sphere $\mathbb{S}^{d-1}_\infty$ at infinity) into the vacuum manifold ${\cal V}$ and may be classified by the corresponding homotopy group $\pi_{d-1} ({\cal V})$. A well-known case occurs when $V$ only depends on the absolute value $\vert \vec \vp \vert$  of $\vec \vp$ and so $V$ obeys $V(\vert \vec \vp \vert =R)=0$ for some $R>0$. The vacuum manifold is then a sphere $\mathbb{S}^{d-1}$ and fields are classified by  the winding number $\pi_{d-1} (\mathbb{S}^{d-1})=\mathbb{Z}$, as is the case, {\it e.g.}, for vortices or monopoles. A further consequence of this is that the corresponding soliton solutions do not take values in the full target space, but instead in the subspace (fundamental region) where $ \vert \vec \vp \vert \le R$ (a $d$-dimensional ball with radius $R$). 

These are the two cases (Skyrme-type or monopole-type topology) which we want to consider in the following. For the function $M$ we assume that it is positive in the whole target space (or, at least, in the fundamental region of the soliton); then $M$ too, has a natural geometrical interpretation. Indeed, let us assume that the target space is equipped with a Riemannian metric
\be \lab{targetspacemetric}
ds^2 = g_{ab} d\vp^a d\vp^b. 
\ee
The correct differential geometric notation requires that the target space coordinate indices are upper indices, {\it i.e.}, $\vp^a$. But it should be obvious that the $\vp_a$ used in the rest of the paper correspond directly to the $\vp^a$ and {\em not} to $g_{ab}\vp^b$. We shall, therefore, return to the notation $\vp_a$ for the target space coordinates to be consistent with the remaining sections (to avoid the possibility of confusion).
The corresponding target space volume form is then
\be
d\Omega^{(d)} = M(\vp_a ) d\vp_1 \wedge \ldots \wedge d\vp_d \qquad \mbox{where} \qquad M \equiv \left( \det (g_{ab}) \right)^{\frac{1}{2}},
\ee
and the pullback of this volume form under the map $\vp_a (\vec x): \IR^d \to {\cal M}$ is
\be
 M\(\vp_a\)\,\sum_{i_1\ldots i_d=1}^d \varepsilon_{i_1\ldots i_d} \partial_{i_1}\vp_1\ldots \partial_{i_d}\vp_d \; dx^1 \wedge \ldots 
\wedge dx^d .
\ee
The first term in the energy functional \rf{actiondagain} (proportional to $M^2$) can therefore be understood as the square of the pullback of the target space volume form, and $M$ is the corresponding volume density. Finally, the topological charge corresponding to the energy \rf{actiondagain} (see \rf{primordialtopcharge}) is given by
\br
Q_{d,M} &=& \int d^d x M\(\vp_a\)\,\sum_{i_1\ldots i_d=1}^d \varepsilon_{i_1\ldots i_d} \partial_{i_1}\vp_1\ldots \partial_{i_d}\vp_d \sqrt{V}
\nonumber \\
&=& k \int_{\cal M'} d^d \vp M(\vp_a) \sqrt{V} = k \int_{\cal M'} d \Omega^{(d)} \sqrt{V} = k {\bf V}({\cal M'}) \langle \sqrt{V}\rangle_{\cal M'},
\er
where ${\cal M'}$ is the fundamental region of the soliton, which coincides with the full target space ${\cal M}$ for skyrmions, but not for vortices or monopoles. Furthermore, $k\in \mathbb{Z}$ is the winding number which takes into account the fact that the soliton $\vec \vp (\vec x)$ may cover the fundamental region ${\cal M'}$ $k$ times while $\vec x$ covers the base space once. The remaining symbols are exactly like in \rf{1d-top-charge}.

\subsection{Symmetries}
For spaces of dimension $d\ge 2$, the energy functional \rf{actiondagain} has a large group of symmetries. Indeed, the antisymmetric combination of derivatives $\sum_{i_1\ldots i_d=1}^d \varepsilon_{i_1\ldots i_d} \partial_{i_1}\vp_1\ldots \partial_{i_d}\vp_d $ in the first term transforms under coordinate transformations $x_i = x_i (y_j)$ with the inverse Jacobian $\det (\frac{\partial y}{\pa x})$ and is, therefore, invariant under coordinate transformations with unit Jacobian, which is volume-preserving diffeomorphisms SDiff$(\mathbb{R}^d)$. In addition, $d^d x$ is invariant under  SDiff$(\mathbb{R}^d)$ by definition, and $M$ and $V$ are scalars. Hence the whole energy functional \rf{actiondagain} is invariant under SDiff$(\mathbb{R}^d)$ coordinate transformations. These are precisely the symmetries of an incompressible fluid and they allow us, therefore, to find new solitons with arbitrary shapes from a given soliton solution with a prescribed ({\it e.g.}, spherically symmetric) shape
\cite{fosco}. 

In addition, as the first term in \rf{actiondagain} can be interpreteed as the square of the pullback of the target space volume form $\d\Omega^{(d)}$, it is obviously invariant under the group of volume-preserving diffeomorphisms on this target space, SDiff$({\cal M})$. The second, potential term in \rf{actiondagain} is, in general, not invariant under the full SDiff$({\cal M})$ group but, depending on its specific form, it may still preserve part of this symmetry. In many cases ({\it e.g.}, the Higgs or Skyrme models), the potential $V$ depends only on the modulus (length) $\vert \vec \vp \vert $ of the fields $\vp_a$, and not on the corresponding ``angular" coordinates. It is, therefore, invariant under SDiff$({\cal M})$ transformations which act nontrivially only on these angular target space variables, but which still form an infinite-dimensional subgroup of SDiff$({\cal M})$.

If the energy functional \rf{actiondagain} for static field configurations is extended to an action in a Lorentz-invariant fashion, like
\be
S = \frac{1}{2} \int dtd^dx \left(  -J^\mu J_\mu - V \right),
\ee
where 
\be
 J^\mu = M(\vp_a ) j^\mu = M(\vp_a )\epsilon^{\mu \mu_1 \cdots \mu_d} \partial_{\mu_1} \varphi_1 \ldots \partial_{\mu_d} \varphi_d 
\ee
then the base space symmetries are reduced to the standard Poincare symmetries, whereas the target space symmetries (the infinite-dimensional subgroup of SDiff$({\cal M})$) survive and are, therefore, promoted to Noether symmetries with the corresponding conservation laws. These theories are, in fact, integrable in the sense of generalised integrability \cite{gen-int} and their conservation laws may be expressed as a generalised zero curvature condition  in an appropriate higher loop space. The fact that, in many cases, the conservation laws of generalised integrability are related to target space SDiff symmetries was first pointed out in \cite{razumov}, for a detailed discussion see \cite{babelon}.

Recently, a specific class of Lorentz non-invariant theories has received considerable interest, namely the so-called Lifshitz type theories \cite{Lifshitz}.  In these theories, characteristically, the kinetic (time-derivative) term is standard (just quadratic in first time derivatives), whereas the space derivative term contains higher than second powers of derivatives, such that the scaling between space and time is inhomogeneous. This, obviously Lorentz symmetry breaking  modification, has the effect of improving the perturbative UV renormalizability of the corresponding QFT while maintaining the standard time evolution (see {\it e.g.} \cite{horava}). In our case, a natural realisation of such a Lifshitz-type theory is achieved by including a non-linear sigma model-type kinetic term based on the target space metric \rf{targetspacemetric} into the action \cite{BPSlifshitz}, leading to
\be
S_{Lif} = \frac{1}{2} \int dtd^dx \left( \sum_{a,b} g_{ab} (\vp_a) \dot \vp_a \dot \vp_b -J^0 J_0 - V \right),
\ee
where $\dot \vp_a = \partial_t \vp_a$. In this case, the base space SDiff$(\mathbb{R}^d)$ symmetries remain intact and are, therefore, Noether symmetries. The 
SDiff$({\cal M})$ group, on the other hand, is broken down to the group of isometries of the target space metric $g_{ab}$.

\subsection{Lower dimensional examples}
Here, let us briefly describe some examples in $d=2$ dimensions. Starting, again, from the energy functional  \rf{actiondagain} and choosing $M=1$ and, {\it e.g.}, the abelian Higgs potential $V_H=(1 - v\bar v )^2$, where $v = \vp_1 + i \vp_2$, we arrive at a kind of ungauged BPS abelian Higgs model which supports vortex-like BPS soliton solutions (these solutions have been computed in \cite{baby-dual}). The deformation method \rf{actiond2} may be used to transform these BPS vortex solutions into BPS vortex solutions for a variety of potentials.  

Next, let us discuss an even simpler transformation. Thus, we start with \rf{actiondagain} and transform it to the new fields $\phi_a$ defined via
the field transformation $\vp_a = \vp_a (\phi_b)$. Then \rf{actiondagain} transforms into
\be
E= \frac{1}{2}\int d^d x\, \left[\( \tilde M\(\phi_a\)\,\sum_{i_1\ldots i_d=1}^d \varepsilon_{i_1\ldots i_d} \partial_{i_1}\phi_1\ldots \partial_{i_d}\phi_d\)^2 + {\tilde V}\(\phi_a\)\right],
\lab{energyd2}
\ee
where
\be 
\tilde M(\phi ) = \vert \frac{\pa \vp}{\pa \phi }\vert M(\vp (\phi )) \; , \quad \tilde V (\phi )= V(\vp (\phi)).
\ee
Note that if we consider the transformations $\vp (\phi)$ that change the boundary conditions, then we may transform between target spaces with different topologies. As a concrete example, let us consider again the BPS vortex model with $M=1$ and $V_H = (1-v\bar v)^2$, together with the field transformation $v\rightarrow u$  ($u = \phi_1 + i \phi_2$) given by
\be \lab{vort-sk-trans}
|v|^2 = (1+|u|^2)^ {-1} \; , \quad \alpha = \beta \qquad \mbox{where} \qquad v=|v|e^{i\alpha} \; , \quad u = |u|e^{i\beta}.
\ee
This transformation maps the fundamental region of the BPS vortex (restricted to $|v|\le 1$) into the full complex plane spanned by $u$. The target space area density $M=1$ (flat space) is transformed into
\be
\tilde M = -(1 + |u|^2)^{-2},
\ee
which (up to a sign) is precisely the area density of the unit two-sphere after a stereographic projection. Finally, the Higgs potential transforms into
\be
\tilde V_H = \left( \frac{|u|^2}{1+|u|^2}\right)^2, 
\ee
which has a unique vacuum at $u=0 \Leftrightarrow \phi_1 = \phi_2 =0$, {\it i.e.}, at the north pole of the target space two-sphere. The transformation \rf{vort-sk-trans} thus transforms BPS vortices characterised by the homotopy group $\pi_1 (\mathbb{S}^1)$ into BPS baby skyrmions with a completely different topology, characterised by the homotopy group $\pi_2 (\mathbb{S}^2)$ (``topological duality" (for more details see \cite{baby-dual})).

%%%%%%%%%%%%%%%%%%%%%%%%%%%%%%%%%%%%%%%%%%%%%
\subsection{An example with $\mathcal{M}=\mathbb{S}^1\times \mathbb{S}^1$}
%%%%%%%%%%%%%%%%%%%%%%%%%%%%%%%%%%%%%%%%%%%%%
Finally,  we consider another choice of the target space geometry which leads to a new type of solitonic solutions which are direct higher-dimensional generalizations of the sine-Gordon solitons. 

In this case we start with two real scalar fields $\phi$ and $\psi$ which are subject to the identification $\phi=\phi +2\pi$ and $\psi = \psi +2\pi$. Hence, the resulting target space is $\mathcal{M}=\mathbb{S}^1\times \mathbb{S}^1$. The volume form on the target space is simply $d\Omega^{(2)}=d\phi \wedge d\psi$ and leads the following static energy
\be
E=\frac{1}{2} \int d^2 x \left( (\epsilon_{ij} \partial_i \phi \partial_j \psi)^2 + V(\phi) W (\psi) \right), 
\ee
where the potential has been assumed to have a factorised form. Furthermore, the ``potentials" $V$ and $W$ should respect our identifications, which simply implies the periodicity in the target space coordinates. Thus we can take, {\it e.g.}, 
\be
V(\phi) = 1-\cos \phi, \;\;\;\;\; W(\psi) = 1-\cos \psi,
\ee
{\it i.e.}, two copies of the Sine-Gordon potential. The resulting Bogomolnyi equation now takes the form
\be
 \epsilon_{ij} \partial_i \phi \partial_j \psi  =\pm  \sqrt{(1-\cos \phi)(1-\cos \psi)} .
\ee 

Its particular solution can be easily found if we further assume that $\phi = \phi(x)$, $\psi = \psi(y)$. Then, the (coupled) Bogomolnyi equation reduces to two (decoupled) Sine-Gordon Bogomolnyi equations
\be
 \phi_x = \sqrt{(1-\cos \phi)}, \;\;\;\  \psi_y = \sqrt{(1-\cos \psi)}
\ee 
with the standard soliton or anti-soliton solutions.  The corresponding topological charge is 
\br
Q &=& \int dxdy  \phi_x \psi_y  \sqrt{(1-\cos \phi)}  \sqrt{(1-\cos \psi)} \\
&=& \int_{-\infty}^\infty dx  \phi_x   \sqrt{(1-\cos \phi)}   \cdot  \int_{-\infty}^\infty dy  \psi_y  \sqrt{(1-\cos \psi)} \\
&\equiv & Q_\phi Q_\psi
\er
which is just the product of two topological charges for two Sine-Gordon theories. Hence, one has two topologically different types of solutions: soliton-soliton (antisoliton-antisoliton) solutions with $Q=1$ and soliton-antisoliton (antisoliton-soliton) with $Q=-1$. 
It is straightforward to extend this construction to any factorized potentials $V(\phi)W(\psi)$ which have at least two vacua in each factor.  

%%%%%%%%%%%%%%%%%%%%%%%%%%%%%%%%%%%%%%%%%%%%%
\section{Examples in three dimensions}
%%%%%%%%%%%%%%%%%%%%%%%%%%%%%%%%%%%%%%%%%%%%%
%%%%%%%%%%%%%%%%%%%%%%%%%%%%%%%%%%%%%%%%%%%%%
\subsection{The BPS Skyrme model}
%%%%%%%%%%%%%%%%%%%%%%%%%%%%%%%%%%%%%%%%%%%%%
\setcounter{equation}{0}
A particular example of a BPS model in three spatial dimensions is provided by the recently discussed BPS Skyrme model defined by the Lagrangian \cite{BPS-Sk}
\be \lab{bps-sk}
{\cal L}=-\lambda^2\,\pi^2 B_{\mu}\,B^{\mu} - \mu^2\, V\(U, U^{\dagger}\)
\ee
with $B^\mu$ being the topological current
\be
B^{\mu}=\frac{1}{24\,\pi^2}\, \varepsilon^{\mu\nu\rho\sigma}\,{\rm Tr}\(L_{\nu}\,L_{\rho}\,L_{\sigma}\),
\qquad\qquad L_{\mu}=U^{-1}\partial_{\mu}U,\qquad\qquad U\in SU(2).
\ee
Equivalently, one can think of $B_\mu$ as being the pullback of the volume form on $\mathcal{M}=\mathbb{S}^3$. 
This model defines a solvable sector of the full Skyrme theory and has found some applications in the context of the low energy limit of QCD \cite{BPS-Sk}, \cite{BPS-Sk-app}. 

Let us begin our discussion here by showing that this model fits into our general framework. 
In the static case what matters is $B_0$, and so
\be
B_0=-\frac{1}{24\,\pi^2}\, \varepsilon_{ijk}\,{\rm Tr}\(L_{i}\,L_{j}\,L_{k}\)=
-\frac{1}{48\,\pi^2}\, \varepsilon_{ijk}\,{\rm Tr}\(L_{i}\,\sbr{L_{j}}{L_{k}}\).
\ee
Denoting by $\vp_i$, $i=1,2,3$, the parameters (fields) of the $SU(2)$ group one gets
\be
L_{i}=U^{-1}\partial_{i} U= \partial_i \vp_a U^{-1}\frac{\delta U}{\delta \vp_a}
=\partial_i \vp_a M_{a b}T_b, 
\ee
where $U^{-1}\frac{\delta U}{\delta \vp_a}=M_{a b}T_b$ is the Maurer-Cartan form of the $SU(2)$ group, and
\be
\sbr{T_a}{T_b}=i\,\varepsilon_{abc}\,T_c, \qquad \qquad \qquad 
{\rm Tr}\(T_a\, T_b\)=\beta\,\delta_{ab}
\ee
with $\beta$ being the Dynkin index of the representation in which the trace is taken. Thus
\br
B_0&=& -\frac{1}{48\,\pi^2}\, \varepsilon_{ijk}
\, \partial_i \vp_{a_1} \, \partial_j \vp_{a_2}\, \partial_k \vp_{a_3}\, M_{a_1 b_1}\, M_{a_2 b_2}\, M_{a_3 b_3}\, {\rm Tr}\(T_{b_1}\,\sbr{T_{b_2}}{T_{b_3}}\)
\nonumber\\
&=&-\frac{i\,\beta}{48\,\pi^2}\, \varepsilon_{ijk}
\, \partial_i \vp_{a_1} \, \partial_j \vp_{a_2}\, \partial_k \vp_{a_3}\, M_{a_1 b_1}\, M_{a_2 b_2}\, M_{a_3 b_3}\, \varepsilon_{b_1 b_2 b_3}.
\er
However, since $a_i, b_i=1,2,3$ we see that 
\be
M_{a_1 b_1}\, M_{a_2 b_2}\, M_{a_3 b_3}\, \varepsilon_{b_1 b_2 b_3}=
\varepsilon_{a_1 a_2 a_3}M_{1 b_1}\, M_{2 b_2}\, M_{3 b_3}\, \varepsilon_{b_1 b_2 b_3}=
\varepsilon_{a_1 a_2 a_3}\,{\rm det}M
\ee
and so
\be
B_0=-\frac{i\,\beta}{48\,\pi^2}\, {\rm det}M\, \varepsilon_{ijk}\,
\, \partial_i \vp_{a_1} \, \partial_j \vp_{a_2}\, \partial_k \vp_{a_3}\, \varepsilon_{a_1 a_2 a_3}
=-\frac{i\,\beta}{8\,\pi^2}\, {\rm det}M\, \varepsilon_{ijk}\,
\, \partial_i \vp_{1} \, \partial_j \vp_{2}\, \partial_k \vp_{3}.
\ee
Thus, $B_0$ has the same form as \rf{topcharged}. 

Note that, in fact,  ${\rm det}M$ is the Haar measure on $SU(2)$. Indeed, the volume element in $SU(2)$ is
\be
dv= \sqrt{{\rm det} \,\eta}\, \, d\vp_1\wedge d\vp_2\wedge d\vp_3
\ee
with
\be
 \eta_{ab}={\rm Tr}\(U^{-1}\frac{\delta U}{\delta\vp_a}\,U^{-1}\frac{\delta U}{\delta\vp_b}\)=M_{ac}M_{bd}{\rm Tr}\(T_c\,T_d\)=\beta\, \(M\,M^T\)_{ab}.
\ee

The relevant topological index (baryon charge) defined by 
\be
q_B =-\int d^3 x \frac{1}{24\,\pi^2}\, \varepsilon_{ijk}\,{\rm Tr}\(L_{i}\,L_{j}\,L_{k}\)
\ee
is just the degree map of one point (at spatial infinity) compactified $\mathbb{R}^3 \cup \{ \infty \} \cong \mathbb{S}^3$ into the target space manifold $SU(2) \cong \mathbb{S}^3$. The usual way of performing  such a compactification is to require that the chiral field $U$ tends to a constant value at the spatial infinity 
\be 
\lim_{|\vec{x}| \rightarrow \infty} U(x)= U_0=const. 
\ee
which can be set to unity by a global transformation. This can be achieved by taking a potential with at least one isolated vacuum at $U_{vac}=1$. The best known example is provided by 
the usual Skyrme potential 
\be V={\rm Tr}\,  (1-U).
  \label{aa}
  \ee

In order to solve the Bogomolny equation it is convenient to use another parametrization of the $SU(2)$ target space. Namely, we put 
\be
U(x)=e^{i\xi(x) \vec{n} \cdot \vec{\tau}},
\ee 
where $\vec{\tau}$ are Pauli matrices, $\xi$ is a real function and $\vec{n}$ is a three component unit vector which spans $S^2$ and so can be related to a complex scalar $u$ by means of the stereographic projection.

In terms of the (target) polar coordinates $\chi$ and $\Phi$ the fields $u$ and $\vec n$ take the form 
\be \lab{u-n-ansatz}
u=\tan \frac{\chi}{2} e^{i\Phi} \; , \quad \vec n = (\sin \chi \cos \Phi ,\sin \chi \sin \Phi , \cos \chi ).
\ee
In these variables the Bogomolny equation takes the form
\be
-\frac{\lambda}{\mu} \frac{\sin^2 \xi}{\sqrt{V}} \sin \chi d\xi d\chi d\Phi = \mp r^2 \sin \theta dr d\theta d\phi,
\label{aaa}
\ee
where we have used spherical coordinates in base space. Equation (\ref{aaa}) can be easily solved by taking
\be
\chi = \theta, \;\;\; \Phi = n\phi
\ee
and
\be 
-\frac{n\lambda}{\mu}  \frac{\sin^2 \xi}{\sqrt{V}} d\xi  = \mp r^2dr.
\ee
In particular, for the usual Skyrme potential (\ref{aa}) one finds that 
\be
\xi = \arccos (\sqrt{2} y-1), \;\;\;\; y \equiv \frac{\mu}{3\sqrt{2} \lambda n} r^3
\ee
for $y \leq \sqrt{2}$ and is zero outside this region. Observe that the map
\be
u: \mathbb{S}^2_{base} \ni (\theta, \phi) \longrightarrow (\chi, \Phi)  \in \mathbb{S}^2_{target}
\ee
is trivially provided by the identification between the base and target space angles while the remaining ``radial" target space coordinate $\xi$ nontrivially depends on the potential. In fact, this way of finding topological solutions can be repeated in higher dimensions. 
%%%%%%%%%%%%%%%%%%%%%%%%%%%%%%%%%%%%%%%%%
\subsection{Monopoles with the SDiff Symmetry}
%%%%%%%%%%%%%%%%%%%%%%%%%%%%%%%%%%%%%%%%%
Let us next discuss a different model. We start by assuming the flat target space manifold $\mathcal{M}=\mathbb{R}^3$ and consider a model based on the following  Lagrangian
\begin{equation}
\mathcal{L}_m= -\frac{\lambda^2}{12^2}\tilde{B}_\mu^2 -\mu^2 \tilde{V}(\vec{\phi}^2), \label{monopole}
\end{equation}
where
\begin{equation}
\tilde{B}^\mu= \epsilon^{\mu \nu \alpha \beta} \epsilon_{abc}\, \phi^a_\nu \phi^b_\alpha \phi^c_\beta.
\lab{phi-top-curr}
\end{equation}
Here, $\vec{\phi}=(\phi^1,\phi^2, \phi^3)$ is a triplet of real scalar fields which span $\mathcal{M}$. Moreover $\phi^i_\mu=\partial_{\mu}\phi^i$.
 In addition we assume that  $\tilde{V}$ is a potential which has a vacuum manifold isomorphic to $\mathbb{S}^2$, that is, $\tilde{V}=0$ if $\vec{\phi}^2=1$. Hence, static, finite energy configurations can be classified by their asymptotic behaviour at spatial infinity $\vec{\phi}_\infty \equiv \lim_{|\vec{x}| \rightarrow \infty} \vec{\phi}: \mathbb{S}^2 \rightarrow \mathbb{S}^2$. The relevant topological index of these maps is the corresponding winding number $k \in \pi_2(\mathbb{S}^2)$.  Such a topological charge is typical for the (3 dim) monopoles and therefore we call objects carrying a nonzero value of this charge ``monopoles". 
Finally, the fundamental region of the monopoles is obviously $|\vec \phi | \le 1$.

To proceed further, as before, we decompose the vector field $\vec \phi$ into a target space radial component and a unit vector  
\begin{equation}
\xi \equiv |\vec{\phi}|, \;\;\; \vec{n}\equiv \frac{\vec{\phi}}{|\vec{\phi}|} \;\;\;\; \mbox{i.e.,} \;\;\;\; \vec{\phi}=\xi \vec{n}.
\end{equation}
Then, 
\begin{equation}
\mathcal{L}_m= -\frac{\lambda^2}{12^2}\tilde{B}_\mu^2 - \mu^2 \tilde{V}(\xi), 
\end{equation}
where now
\begin{equation}
\tilde{B}^\mu=3\xi^2 \epsilon^{\mu \nu \alpha \beta} \epsilon_{abc}\, \xi_\nu n^a n^b_\alpha n^c_\beta \equiv 3\xi^2 \epsilon^{\mu \nu \alpha \beta} \xi_\nu H_{\alpha \beta}
\end{equation}
and 
\begin{equation}
H_{\alpha \beta} \equiv \epsilon_{abc}\, \xi_\nu n^a n^b_\alpha n^c_\beta = 2i \frac{u_\alpha \bar{u}_\beta-u_\beta \bar{u}_\alpha}{(1+|u|^2)^2}.
\end{equation}
Here and  in what follows, we have used the notation that a subscript $\mu$ when attached to $n^i$, $\xi$, $u$ and $\bar u$ denotes the derivative of this quantity (like just above for $\phi^i$ in \rf{phi-top-curr}).

The last step above has involved the stereographic projection from $\vec n$ to $u$. Hence, the fields of the model may be expressed by one real and one complex field and the Lagrangian takes the form:
\begin{equation}
\mathcal{L}_m =-\frac{\lambda^2\xi^2}{(1+|u|^2)^4} \left(  \epsilon^{\mu \nu \alpha \beta} \xi_\nu u_\alpha \bar{u}_\beta \right)^2  - \mu^2 \tilde{V}(\xi)
\end{equation}
with the vacuum located at $\xi=1$ (or in general at a non-zero vale of the real scalar).

Assuming again the spherical ansatz \rf{u-n-ansatz} we find the following solution of the Bogomolnyi equation:
\begin{equation}
\chi = \theta, \;\;\;\; \Phi =n\phi
\end{equation}
while the real scalar $\xi$ satisfies
\begin{equation}
\frac{n^2\lambda^2}{4\mu^2r^4} \xi^2\xi_r^2=\tilde{V}
\end{equation}
or 
\begin{equation}
\frac{1}{4} \xi^2\xi_z^2=V \;\; \Rightarrow \;\; \frac{1}{2} \xi\xi_z=\pm \sqrt{\tilde{V}},
\label{v77}
\end{equation}
where a new variable $z$ has been defined by $z=\frac{4\mu r^3}{3\lambda |n|}$. The equation (\ref{v77}) must be solved with appropriate boundary conditions providing nontrivial topology and regularity {\it i.e.}, $\xi(0)=0$ and $\xi(Z)=1$, where $Z$ can be finite (compactons) or infinite (infinitely extended solitons).

To solve (\ref{v77})  we have to specify the potential which is compatible with the previous requirements.  For example, we can take a generalized Higgs potential 
\begin{equation}
\tilde{V}=(\vec{\phi}^2 -1)^{2a} = (\xi^2 -1)^{2a}.
\end{equation}
Then, the obvious solutions of (\ref{v77}) are:
\\
i) compactons for $a \in (0,1)$
\begin{equation}
\xi = \left\{
\begin{array}{cc}
\sqrt{1-\left( 1-\frac{z}{Z}\right)^{\frac{1}{1-a}}} & z \in [0,Z],
\\
&
\\
1 & z \geq Z \equiv \frac{1}{1-a},
\end{array}
\right.
\end{equation}
where $Z$ is the compacton radius;
\\
ii) an exponentially localized solution for $a=1$
\begin{equation}
\xi=\sqrt{1-e^{-z}},
\end{equation}
iii) and power-like localized solutions for $a>1$
\begin{equation}
\xi=\sqrt{1-\left(\frac{a-1}{a-1+z}\right)^{\frac{1}{a-1}}}.
\end{equation}
%%%%%%%%%%%%%%%%%%%%%%%%%%%%%%%%%%%%%%%%%
\subsection{Target space transformation as topological duality}
%%%%%%%%%%%%%%%%%%%%%%%%%%%%%%%%%%%%%%%%%
Let us now return to the BPS Skyrme model given in \rf{bps-sk}. Note that it differs from the previously introduced BPS monopole model only by the $\xi$-dependent factor multiplying the first term, {\it i.e.}, by a different target space volume density.  Obviously, this is a reflection of  different target space metrics corresponding to these models.  Moreover, in the BPS Skyrme models we should also assume that the potential has isolated minima in the target space $SU(2)\cong \mathbb{S}^3$. The relevant topological charge is now the baryon charge which requires the following boundary conditions for the chiral field
\begin{equation}
\lim_{|\vec{x}| \rightarrow \infty} U=1 \;\; \Rightarrow \;\; \lim_{|\vec{x}|\rightarrow \infty} \xi = 0. 
\end{equation}
Thus, in this case, static solutions can be treated as maps from the compactified three dimensional Euclidean base space into the target space manifold {\it i.e.} $U: \mathbb{R}^3\cup \{ \infty \} \cong \mathbb{S}^3 \; \rightarrow \;  SU(2) \cong \mathbb{S}^3$. Hence, $q_B \in \pi_3(\mathbb{S}^3)$. In order to satisfy this boundary condition the potential must have its vacuum at $\xi =0$.

Note that we have a ``topological duality" map between a solution $\vec{\phi}$ of the BPS monopole model and a solution $U$ of the BPS Skyrme model. This map is provided by a transformation between their $\xi$ fields,  namely, 
\begin{equation}
\xi^2_m = \frac{1}{\pi} \left( \pi- \xi_s+\cos \xi_s \sin \xi_s \right), \;\;\;\; u_m=u_s \qquad \mbox{(the}\quad \mbox{same)},
\label{v83}
\end{equation}
where $(\xi_m, u_m)$ are functions defining a BPS monopole $\vec{\phi}$, while $(\xi_s, u_s)$ parametrize a BPS skyrmion. One can check that the boundary conditions necessary for the existence of monopoles ($\xi_m(0)=0, \; \xi_m(\infty)=1$) are transformed  by (\ref{v83}) into the skyrmion boundary conditions ($\xi_s(0)=\pi, \; \xi(\infty)=0$). The transformation, therefore, maps the fundamental region $|\vec \phi| \le 1$ of the monopole into the full target space $S^3$ of the skyrmion.
\\
This  map fully connects both models (the corresponding field equations) if the potentials obey
\begin{equation}
\tilde{V}(\xi_m)=V(\xi_s).
\end{equation}
Hence, the BPS Skyrme models dual to the BPS monopole model with the generalized Higgs potential have the following potential term
\begin{equation}
V=\left(\frac{\xi-\cos \xi \sin \xi}{\pi} \right)^{2a}
\end{equation}
which, in fact,  has been recently discussed in the context of the BPS Skyrme model coupled with the vector mesons and is referred to as the BPS potential.
\\
On the other hand, the monopole model dual to the BPS Skyrme model with the usual potential $$V=(1-\cos \xi)$$ cannot be written in a simple form, as the duality map is quite complicated if one wants  to express it as $\xi_s=F(\xi_m)$. Nonetheless, such a dual formulation exists and, in principle, the usual BPS skyrmions can be mapped, in a one-to-one way, into the BPS monopoles with the $\pi_2(\mathbb{S}^2)$ topology. 
\\
Obviously, this map is a higher dimensional generalization of the non-holomorphic ``topological duality"  map recently observed for the BPS baby skyrmions and the BPS vortices and briefly discussed in Section IV.D. In both cases, the duality maps transform the profile function of one model into the profile function of the other, while the ``angular' field variables (which fix the topological charges of the solutions) remain unchanged. 

%%%%%%%%%%%%%%%
\section{Conclusions}
%%%%%%%%%%%%%%%%
In this paper, we have introduced an appropriately generalised concept of self-duality as the basic tool for the construction and analysis of a large class of BPS theories in arbitrary dimensions. In (1+1) dimensions, this enabled us to re-derive and interpret, in a simple fashion, the deformation procedure of Bazeia and collaborators, which permits the construction of infinitely large families of theories supporting kink solutions from given ``seed" configurations (static solutions of other theories). 

Furthermore, our approach to self-duality has allowed us to generalise
to higher dimensions both the construction of BPS theories and the deformation procedure in a relatively simple manner. Specifically, for a class of higher-dimensional scalar field theories, whose topology is provided by a volume form on the corresponding target space, we have demonstrated that topological solitons with different topologies may be transformed into each other by simple field transformations. We have also given several explicit examples of field theories and their BPS soliton solutions in two and three dimensions, like, {\em e.g.}, the BPS Skyrme model (which has already found some applications as an effective low energy field theory for strong interaction physics), and a BPS monopole model related to the former model by a field transformation. 

One question of interest which deserves further study involves the investigation of whether further models of physical or phenomenological relevance ({\em e.g.}, in cosmology), may be found among the classes of theories investigated in the present paper. Finally, we would like to remark that the notion of self-duality introduced in Section II applies to a much larger class of BPS theories than the ones considered here. Another interesting question is, therefore, whether this generalised self-duality framework may be helpful in the construction and analysis of new BPS models not considered so far. These issues are currently under investigation.

\vspace*{0.3cm}

{\centerline {\bf Acknowledgement}}

\vspace*{0.2cm}

CA and AW acknowledge financial support from the Ministry of Education, Culture and Sports, Spain (grant FPA2008-01177), the Xunta de Galicia (grant INCITE09.296.035PR and Conselleria de Educacion), the Spanish Consolider-Ingenio 2010 Programme CPAN (CSD2007-00042), and FEDER. 
Further, AW was supported by Polish NCN grant 2011/01/B/ST2/00464.
LAF is partially supported by CNPq-Brazil. 
The authors are grateful  to J. S\'anchez Guill\'en for fruitful
discussions. 
Some part of the work reported in this paper was
carried out when WJZ visited AW in Cracow.
He would like to thank AW for the invitation and the Jagiellonian University
for the hospitality.


\begin{thebibliography}{99}

\bibitem{man-sut-book}
N. Manton, P. Sutcliffe, "Topological Solitons", Cambridge University Press, Cambridge, 2007.
\bibitem{bogo} E.B. Bogomolnyi, ````The stability of Classical Solutions''
{\it Sov. J. Nucl. Phys.} {\bf 24} 449, 1976.
\bibitem{prasad} M. K. Prasad, C. M. Sommerfield, Phys. Rev. Lett. {\bf 35} (1975) 760.

\bibitem{bazeia}
D.~Bazeia, L.~Losano, J.~M.~C.~Malbouisson,
Phys. Rev. D{\bf 66} (2002) 101701;
 C. A. Almeida, D.~Bazeia, L.~Losano, J.~M.~C.~Malbouisson,
Phys.Rev. D69 (2004) 067702;
D.~Bazeia, L.~Losano, J.~M.~C.~Malbouisson, R.~Menezes,
  Physica D{\bf 237}, 937 (2008).

\bibitem{tch} H. J. W. Muller-Kirsten and D. H. Tchrakian, J. Phys. {\bf A23} (1990) L363; Phys. Rev. {\bf D44} (1991) 1204.
\bibitem{GiPa}
 T. Gisiger and M.B. Paranjape, Phys. Rev. D{\bf 55} (1997) 7731.
\bibitem{BPS-bS} C. Adam, T. Romanczukiewicz,  J. Sanchez-Guillen, A. Wereszczynski, 
Phys. Rev. D{\bf 81} (2010) 085007.
\bibitem{Sp1}
J.M. Speight, 
J. Phys. A{\bf 43} (2010) 405201.  
\bibitem{BPS-Sk}
C. Adam, J. Sanchez-Guillen, A. Wereszczynski,
Phys. Lett. B{\bf 691}, 105 (2010)
[arXiv:1001.4544];
Phys. Rev. D{\bf 82}, 085015 (2010)  
[arXiv:1007.1567].
\bibitem{fosco} C. Adam, C.D. Fosco, J.M. Queiruga, J. Sanchez-Guillen, A. Wereszczynski, J. Phys. {\bf A46} (2013) 135401; [arXiv:1210.7839].
\bibitem{baby-dual} 
C. Adam, J. Sanchez-Guillen, A. Wereszczynski, W. J. Zakrzewski, Phys.Rev. D{\bf 87} (2013) 027703 [ arXiv:1209.5403]. 

\bibitem{skyrme} T.H.R. Skyrme, Proc. Roy. Soc. Lon. {\bf 260},
127 (1961); Nucl. Phys. {\bf 31}, 556 (1962); J. Math. Phys. {\bf
12}, 1735 (1971).

\bibitem{baby} B.M.A.G. Piette, B.J. Schoers and W.J.
Zakrzewski, Z. Phys. C {\bf 65} (1995) 165; B.M.A.G. Piette, B.J. 
Schoers and W.J. Zakrzewski, Nucl. Phys. B {\bf 439} (1995) 205.
\bibitem{holom} R.A. Leese, M. Peyrard and W.J. Zakrzewski
Nonlinearity {\bf 3} (1990) 773; B.M.A.G. Piette and W.J.
Zakrzewski, Chaos, Solitons and Fractals {\bf 5} (1995) 2495; P.M. 
Sutcliffe, Nonlinearity (1991) {\bf 4} 1109.

\bibitem{BPS-Sk-app}
E. Bonenfant, L. Marleau,
Phys. Rev. D{\bf 82}, 054023 (2010)
[arXiv:1007.1396];  E. Bonenfant, L. Harbour, L. Marleau, Phys. Rev. D{\bf 85} (2012) 114045.

\bibitem{fer-zak}
L.A. Ferreira, W.J. Zakrzewski, JHEP {\bf 1105} (2011) 130; JHEP {\bf 1209} (2012) 103; Int. J. Geom. Meth. Mod. Phys. {\bf 09} (2012) 1261004.

\bibitem{bazeia2}
D. Bazeia, L. Losano, J.R.L. Santos,   arXiv:1304.6904.

\bibitem{casana}
L. Losano, J.M.C. Malbouisson, D. Rubiera-Garcia, C. dos Santos, Eur.
Phys. Lett. {\bf 101} (2013) 31001; R. Casana, M.M. Ferreira, E. da Hora,
C. dos Santos, Phys.Lett. B{\bf 722} (2013) 193; L. Sourrouille, Phys. Rev.
D{\bf 87}, 067701 (2013); R. Casana, M.M. Ferreira, Jr, E. da Hora,
Phys.Rev. D{\bf 86} (2012) 085034; D. Bazeia, R. Casana, E. da Hora, R.
Menezes, Phys.Rev. D{\bf 85} (2012) 12502.  

\bibitem{gen-int} O. Alvarez, L.A. Ferreira, J. Sanchez-Guillen, Nucl. Phys.
{\bf B529} (1998) 689; Int. J. Mod. Phys. A {\bf 24}, 1825 (2009).

\bibitem{razumov}
L.~A.~Ferreira and A.~V.~Razumov, Lett. Math. Phys. {\bf 55},  (2001) 143
[hep-th/0012176].

\bibitem{babelon}
O. Babelon, L.~A. Ferreira,
 JHEP {\bf 0211} (2002) 020 
[hep-th/0210154]. 

\bibitem{Lifshitz} E. Lifshitz Zh.Eksp Teor. Fiz. {\bf 11} (1941) 255.
\bibitem{horava} P. Horava, Phys. Lett B{\bf 694} (2010) 172;
 Phys. Rev. D{\bf 79} 084008 (2009).
\bibitem{BPSlifshitz}
C. Adam, C. Naya, J. Sanchez-Guillen, A. Wereszczynski,
JHEP {\bf 1303} (2013) 012. 
 
 \end{thebibliography}
 \end{document}